\documentclass[%
 reprint,
 amsmath,amssymb,
 aps,
]{revtex4-2}

\usepackage{graphicx}
\usepackage{dcolumn}
\usepackage{bm}

\usepackage{graphicx}
\usepackage{dcolumn}
\usepackage{bm}
\usepackage{longtable}
\usepackage{amssymb}
\usepackage{xcolor}
\usepackage{hyperref}
\hypersetup{colorlinks=true,breaklinks,linkcolor=blue,urlcolor=blue,citecolor=blue}

\begin{document}

\preprint{APS/123-QED}

\title{Origin of phase stability in Fe with long-period stacking order as an intermediate phase in cyclic $\gamma$--$\epsilon$ martensitic transformation} 

\author{Takao Tsumuraya$^{1, 2}$}\thanks{tsumu@kumamoto-u.ac.jp}
\affiliation{$^{1}$International Center for Young Scientists, National Institute for Materials Science,  Tsukuba 305-0044, Japan}
\affiliation{$^{2}$Priority Organization for Innovation and Excellence, Kumamoto University, Kumamoto 860-8555, Japan}
\author{Ikumu Watanabe$^{3}$ and Takahiro Sawaguchi$^{3}$}\thanks{sawaguchi.takahiro@nims.go.jp}
\affiliation{$^{3}$Research Center for Structural Materials, National Institute for Materials Science, Tsukuba 305-0047, Japan}

\date{\today}

\begin{abstract}
A class of Fe-Mn-Si-based alloys exhibits a reversible martensitic transformation between the $\gamma$ phase with a face-centered cubic~(fcc) structure and an $\epsilon$ phase with a hexagonal close-packed (hcp) structure. 
During the deformation-induced $\gamma$--$\epsilon$ transformation, we identified a phase that is different from the $\epsilon$ phase. In this new phase, the electron diffraction spots are located at 
the 1/3 positions that correspond to the $\{$0002$\}$ plane of the $\epsilon$ (hcp) phase with 2H structure, which suggests long-period stacking order (LPSO). 
To understand the stacking pattern and explore the possible existence of an LPSO phase as an intermediate between the $\gamma$ and $\epsilon$ phases,
the phase stability of various structural polytypes of iron was examined using first-principles calculations with a spin-polarized form of the generalized gradient approximation in density functional theory.
We found that an antiferromagnetic ordered 6H$_2$ structure is the most stable among the candidate LPSO structures and is energetically close to the $\epsilon$ phase, which suggests that the observed LPSO-like phase adopts the 6H$_2$ structure. 
Furthermore, we determined that the phase stability can be attributed to the valley depth in the density of states, close to the Fermi level. 
\end{abstract}

\pacs{Valid PACS appear here}
\maketitle

\section{Introduction}
\label{intro}
Austenitic steel is an industrial structural material with a long history. 
Ever since wear-resistant Fe-Mn-C steel was developed at the end of the 19th century, much attention has been paid to its superior mechanical properties.
Among these alloys, those with 28--32 mass percent (mass$\%$) Mn and 4--7 mass$\%$ Si are known to demonstrate a shape-memory effect. 
This effect is governed by a non-diffusive solid-to-solid phase transformation from $\gamma$-austenite with a face-centered cubic~(fcc) structure~to $\epsilon $-martensite with a hexagonal close-packed (hcp) structure~\cite{Sato_FeMnSi_1986}.
Plastic deformation and subsequent shape recovery upon heating are associated with the forward $\gamma $ $\rightarrow$ $\epsilon$ and reverse $\epsilon$ $\rightarrow$ $\gamma$ transformations, respectively~\cite{Ogawa_Kajiwara93_FeMnSi, Kikuchi_Kajiwara_Tomota_FeMnSi, ohtsuka1995growth, SAWAGUCHI_2006}.

\begin{figure}[tb]
\begin{center}
\includegraphics[width=0.8\linewidth]{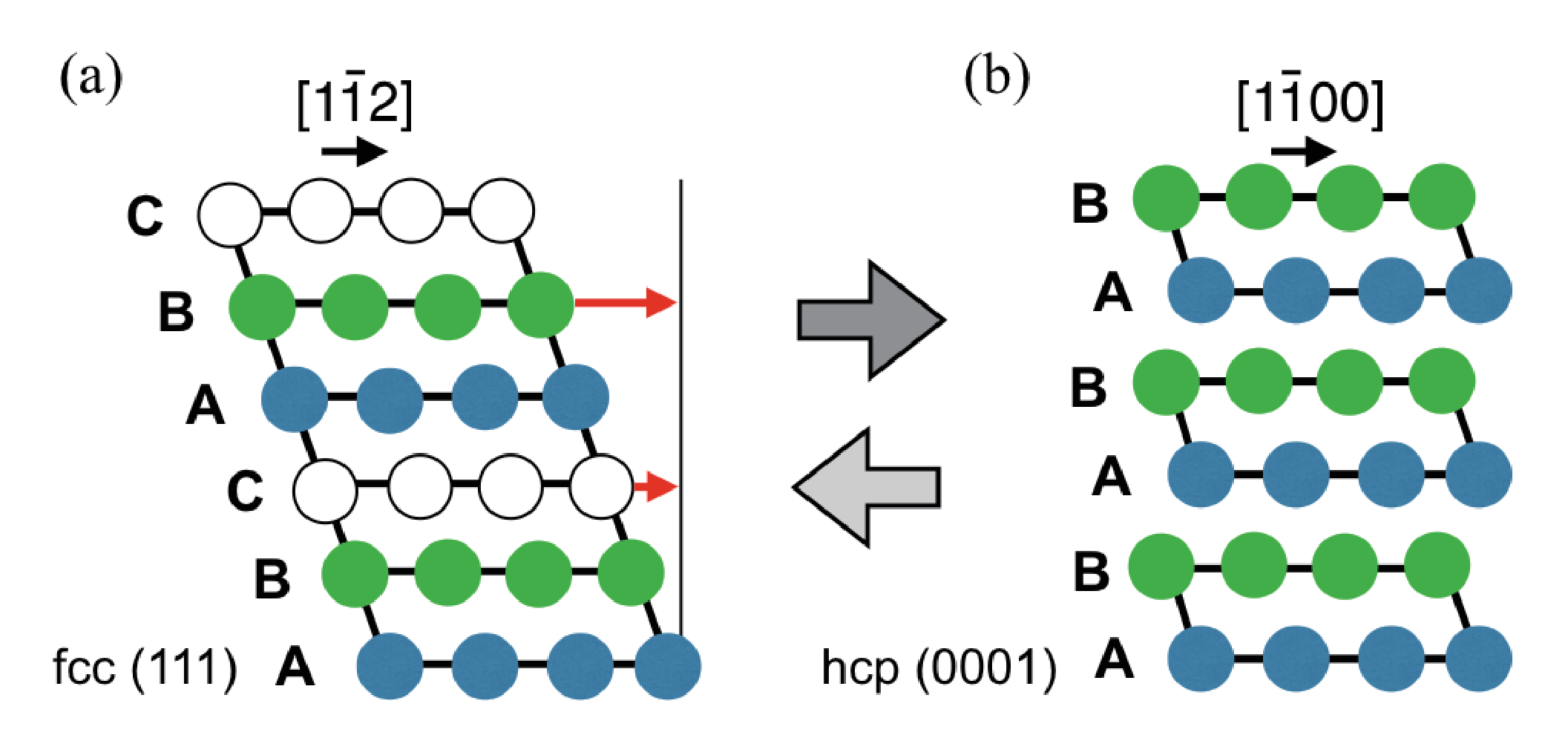}
\end{center}
\setlength\abovecaptionskip{-3pt}
\caption{Atomic displacement and formation of partial dislocation-stacking
fault units associated with (a)~$\gamma$~$\rightarrow$~(b)~$\epsilon$ martensitic transformation.}
\label{gamma_epsilon}
\end{figure}

Figure~\ref{gamma_epsilon} illustrates the atomic displacement during the $\gamma$ $\leftrightarrow$ $\epsilon$ transformation. 
The (111) planes in the fcc structure are parallel to the basal (0001) planes of the hcp lattice.
The formation of the $\epsilon$ phase from the $ \gamma$ phase is induced by stacking faults bounded by Shockley partial dislocations with an $a$/$\sqrt{6}$ shift on the (111) plane. 
These partial dislocations occur every two layers in the pathway from the fcc to the hcp structure~\cite{nishiyama1978martensitic}.
This transformation is one of the notable plastic deformation modes in austenite steels.  
The reversible transformation between the $\gamma$ and $\epsilon$ phases occurs during heating and cooling cycles, and cyclic plastic deformation.  

Since researchers showed that a dual-phase magnesium-based alloy with a long-period stacking order (LPSO) structure and $\alpha$-Mg (hcp) exhibited superior mechanical properties and a tensile yield strength of approximately 600 MPa~\cite{kawamura2001rapidly},
there has been a growth in the development of novel alloys with LPSO structures~\cite{abe2002long, LPSO_TiAl2020}.
In the 1960s, Lysak and Nikolin discovered a phase 
in a disordered Fe-Mn-C alloy subjected to heating and cooling~cycles of 400 $\leftrightarrows$ --196~$^\circ$C that was distinct from the $\epsilon$ (hcp) phase~\cite{Lysak_1963, Lysak_1967, Kononenko_FeMn_1969, Lysak1970effect}.
The new phase was referred to as the $\epsilon^{\prime}$ phase.
Many LPSO-like phases were later discovered in various Fe-Mn-(Al)-C-based alloys~\cite{Oka_LPSO_FeMnC_72, Oka_FeMnC1973, Lysak_1967, ohtsuka1995growth, HWANG_Fe_LPSO_1991, Chao_18R_Marten1992, LEE199521}. 
Although the existence of LPSO phases in Fe-Mn-Si-based alloys has also been verified by thermodynamic modeling~\cite{Wan2001_Thermo_FeMnSi}, 
the actual stacking pattern of LPSO phases has yet to be experimentally identified.
In addition, even in pure iron, the phase stability and magnetic properties of the structural polytypes have not yet been studied using first-principles calculations.

In the present work, transmission electron microscopy (TEM) measurements were performed on Fe-Mn-Si-based alloys subjected to cyclic deformation. 
Electron diffraction spots at the 1/3 positions that correspond to the $\{$0002$\}$ plane of the $\epsilon$ (hcp) phase with 2H structure suggest the existence of an LPSO structure.
However, for Fe-Mn-Si-based alloys under cyclic deformation, the observed LPSO phase was unstable during room temperature aging (probably due to sensitivity to temperature variations), and so the structural and magnetic properties of the phase are still unavailable.  
Therefore, we aimed to determine the most stable stacking configuration of the LPSO structure. 
The structural and magnetic phase stability of structural polytypes in pure iron were investigated using first-principles calculations based on density functional theory (DFT)~\cite{Hohenberg_Kohn, Kohn_Sham}. 
The relative stabilities of the fcc and hcp phases were examined by structural optimization to understand the possible realization of an LPSO phase as an intermediate phase in the~$\gamma$--$\epsilon$ transformation. 

Antiferromagnetic~(AFM) order is crucial in the stabilization of Fe--Mn-based alloy phases with fcc and hcp structures ~\cite{Ohno_AFM_hcpFeMn, Endoh_FeMn, Kennedy_1987, Bleskov_FeMn_MagSFE2016, ReyesFeMn}.
Therefore, we consider all the possible AFM spin structures with collinear spin order, and compare the total energies obtained from non-spin-polarized calculations.
The total energies of the candidate LPSO structures are compared to the more stable structure of hcp Fe at 0 K. We discuss the origin of structural stabilities based on the difference in the total density of states (DOS) near the Fermi level.

For the sake of simplicity, pure iron is used as a model for those Fe-rich Fe-$X$ alloys; the amount of alloying element ($X$) is relatively low ($\leq$ 25 atom$\%$), and the Fe content is large.
The Fe ground state at ambient pressure is widely known to be ferromagnetic with a body-centered cubic ($bcc$) structure; hcp Fe with a nonmagnetic (NM) ground state only appears under high pressure~\cite{Taylor_Fe_Mossb82, Nicol_ScienceFe}.
In contrast, various hexagonal Fe-rich Fe-$X$ alloys exhibit a wide variety of magnetic states at both ambient and high pressures.
Fe-rich Fe--Mn-based alloys with the hcp structure ($\epsilon$ phase) exhibit a N{\'e}el temperature of 230~K~\cite{OhnoH_Ru_Os71}.
$\epsilon$--Fe-$X$ alloys ($X$ = Ru, and Os) have AFM ground states at ambient pressure with N{\'e}el temperatures of approximately~100~K. 
The Fe$_2$Ta alloy exhibits a paramagnetic state in which either an excess of Fe or Ta can induce ferromagnetic ordering at low temperatures (approximately~150~K)~\cite{Kai_Fe-Ta1970, SergiuFeTaPRB}.
The study of structural polytypes of pure iron will also facilitate investigations into the possibility of metastable phases in such Fe-rich alloys and the new $\epsilon$ phase found in pure iron under extremely high pressures~\cite{Saxena1703, Yoo1996_dhcp_Fe_P}.

\begin{figure}[tb]
\begin{center}
\includegraphics[width=0.85\linewidth]{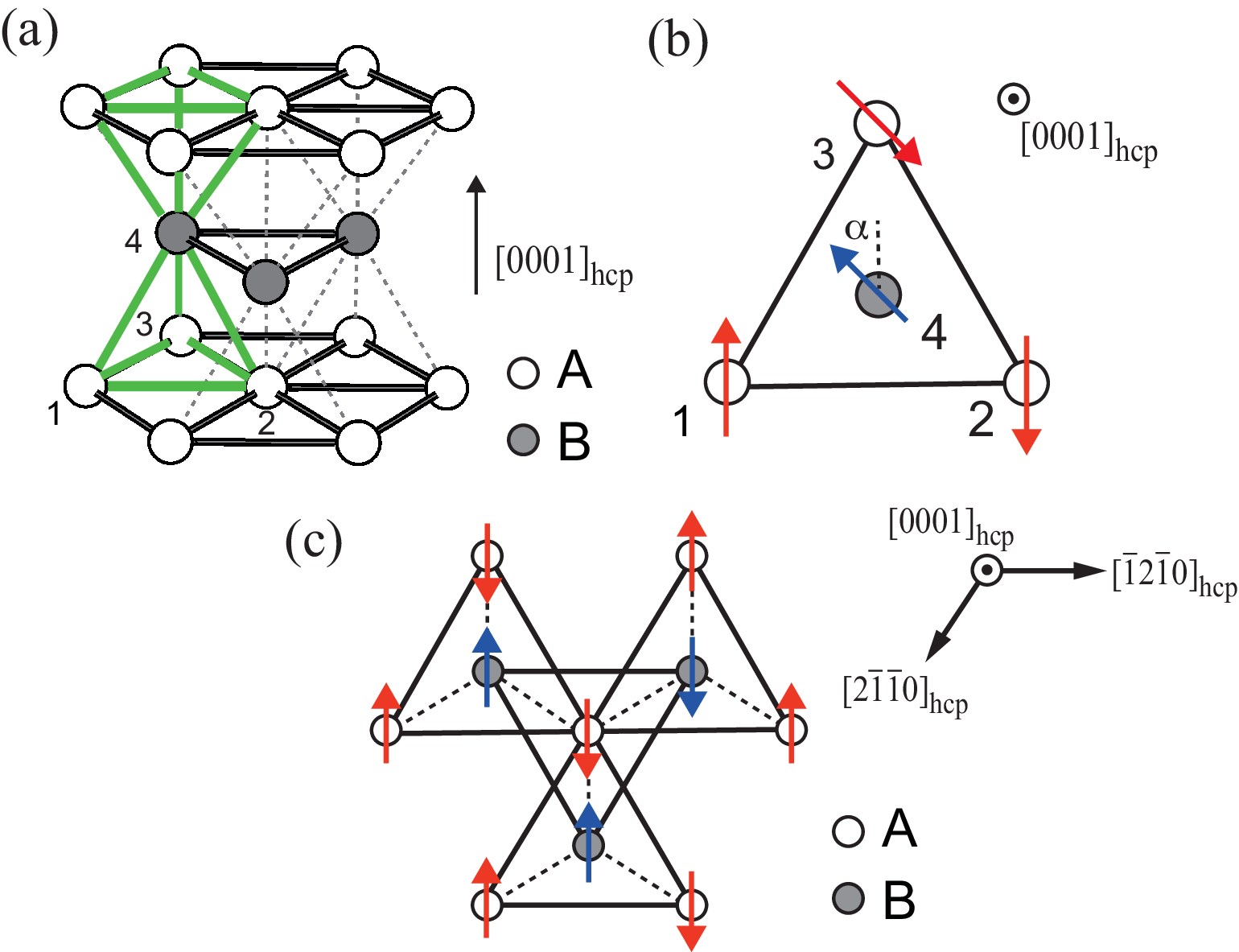}
\end{center}
\setlength\abovecaptionskip{-5pt}
\caption{(a) Atomic configuration of the hcp lattice. 
The tetrahedron marked in bold green lines is the spin frustration unit in the hcp lattice. 
(b) An example of spin frustration in a tetrahedron into which an hcp lattice can be decomposed.  
Arrows show the spin orientation on an isolated tetrahedron in the hcp lattice. 
(c) Spin configuration of the AFM-II phase in hcp-Fe, which is projected on the (0001)$_{hcp}$ plane. 
Up and down arrows indicate the spin orientation. 
The solid line connects in-plane sites, and
the dashed line connects sites to those in the adjacent layer.  
The open circles with red arrows show the internal atomic coordinates at $z$ = 1/4, and the filled circles with blue arrows show those at $z$ = 3/4.} 
\label{hcp_tetra}
\end{figure}

It is also notable that determination of the ground state for both fcc and hcp Fe from first-principles calculations has proved challenging because the energy difference between different magnetic ordering patterns is constrained to a narrow energy window~\cite{Asada_Terakura_Fe1992, Jiang_Carter_fcc_Fe2003, Thakor_Fe_finite_temp03, Cohen2004, Lizarraga_Fe_PRB08, Fe_nematic_Lebert2019}.
Magnetic frustration occurs in hcp Fe, where spins are expected to be geometrically frustrated with respect to AFM order~\cite{Auerbach_PRB_88, Diep_hcp_frust92}. 
Each spin is shared by the eight tetrahedra of the hcp lattice, and the nearest neighbor bonds are shared by the two tetrahedra, as shown in Fig.~\ref{hcp_tetra} (a). 
The ground state spin configuration of the system is formed by the stacking of adjacent tetrahedra.
The following two choices are present. 
Suppose that two spins are antiparallel (between 1 and 2) [Fig.~\ref{hcp_tetra}(b)], 
in which case, the two remaining spins (between 3 and 4) should also be antiparallel. 
However, these axes can be chosen to form an arbitrary angle (an infinite number of ways) with respect to the axis of the first two spins; 
the ground state is thus infinitely degenerate.
If the first two spins form an angle $\alpha$, then the other two spins must form the same angle and be antiparallel with the first two spins~\cite{Diep_hcp_frust92}. 

As a result of the geometrical frustration of the spin moments arranged in a tetrahedral configuration in hcp Fe, noncollinear spin order ($\alpha$ $\neq$ 0) and spin-intensity modulated (spin-smectic) phases arise as local minima or saddle points by consideration of an isolated tetrahedron~\cite{Thakor_Fe_finite_temp03, Cohen2004, Lizarraga_Fe_PRB08, Fe_nematic_Lebert2019}. 
However, previous DFT studies predicted that a collinear AFM state known as type II (AFM-II) is the lowest energy spin configuration of $\epsilon$ Fe at ambient pressure~\cite{Steinle-Neumann_Fe99, Steinle-Neumann_Fe_Eratum, Fe_nematic_Lebert2019}. 
This collinear spin structure is represented by $\alpha$ = 0 \cite{Cohen2004}, and each atom has eight antiferromagnetically coupled and four ferromagnetically coupled neighbors, as illustrated in Fig.~\ref{hcp_tetra}(c). 
Their calculated bulk moduli and lattice parameters show better agreement with the recent experimental equation of state (EoS)~\cite{Dewaele2006, YamazakiEoSFe2012, Sakai2014, Fei_Murphy2016}, compared to the values obtained from NM calculations.
In this study, we have systematically searched for collinear AFM ordering of various LPSO phases and compare the stability over the total energy of the hcp AFM-II phase.

The paper is organized as follows. 
The details of the first-principles DFT method are given in Section~\ref{Calc}. 
Experimental observations of LPSO-like phases in Fe-Mn-Si-based alloys are presented in Section~\ref{TEM}.
Section~\ref{crystals} describes the calculated structural models for the structural polytypes of Fe. 
The overall procedure for exploration of the stable AFM patterns is given in Section~\ref{strucAFM}.
The calculated structural and magnetic stabilities are discussed using the EoS in Section~\ref{DFTcalc}. 
Finally, Section~\ref{origin} presents an analysis of the DOS for AFM states of LSPO structures to discuss the electronic origin of phase stabilities, followed by our conclusions.  

\section{Calculation methods}
\label{Calc}
First-principles DFT calculations were performed using the all-electron full-potential linearized augmented plane wave (FLAPW) method implemented in the \textsf{QMD-FLAPW12} code~\cite{Wimmer1981, Weinert, L_KA}.
This method is known as the most accurate among the first-principles methods.
The exchange-correlation functional used was a spin-polarized form of the generalized gradient approximation (GGA) proposed by Perdew, Burke, and Ernzerhof (PBE)~\cite{GGA_PBE}.
For NM phases, integration of the Brillouin zone was performed using $\bm{k}$-point grids of 16$\times$16$\times$8 for hcp (2H) Fe, 12$\times$12$\times$12 for fcc (3C), 16$\times$16$\times$4 for dhcp Fe, and 16$\times$16$\times$4 for 6H$_1$, 6H$_2$, and 10H structures. 
The symbols $n$H and 3C represent structural polytypes in Ramsdell notation, where $n$ refers to the stacking period (total number of close-packed planes in the unit cell), and the letters H and C denote the hexagonal and cubic lattice types. The number 3 in 3C refers to the stacking period in close-packed layers (\textsf{ABC}), and 3C is the only possible cubic polytype.
The subscript in the 6H structures specifies a different stacking configuration of the close-packed planes. 
The $\bm{k}$--point mesh used for orthorhombic cells with an AFM order of 2H, 6H$_1$ and 6H$_2$ is 10$\times$6$\times$4, while the meshes for tetragonal unit cells with AFM-S and AFM-D states in fcc (3C) structures are 12$\times$12$\times$8 and 12$\times$12$\times$4, respectively. 
To accurately obtain the electronic structure and the EoS, the cut-off energies for the LAPW basis functions in the interstitial were set to 36 and 310 Ry for the plane waves and potentials, respectively. 
The common muffin-tin (MT) sphere radius of Fe was set to 1.16~\AA~for all structural polytypes of Fe. 
The angular momentum expansion inside the MT sphere was truncated at $l$ = 8 for Fe atoms. 
Finally, the method of explicit orthogonalization was used in the present study~\cite{Weinert_2009}. 

\section{Experiment}
\label{TEM}
\subsection{Sample preparation and set-up for fatigue test}
\begin{figure}[b]
\begin{center}
\includegraphics[width=0.85\linewidth]{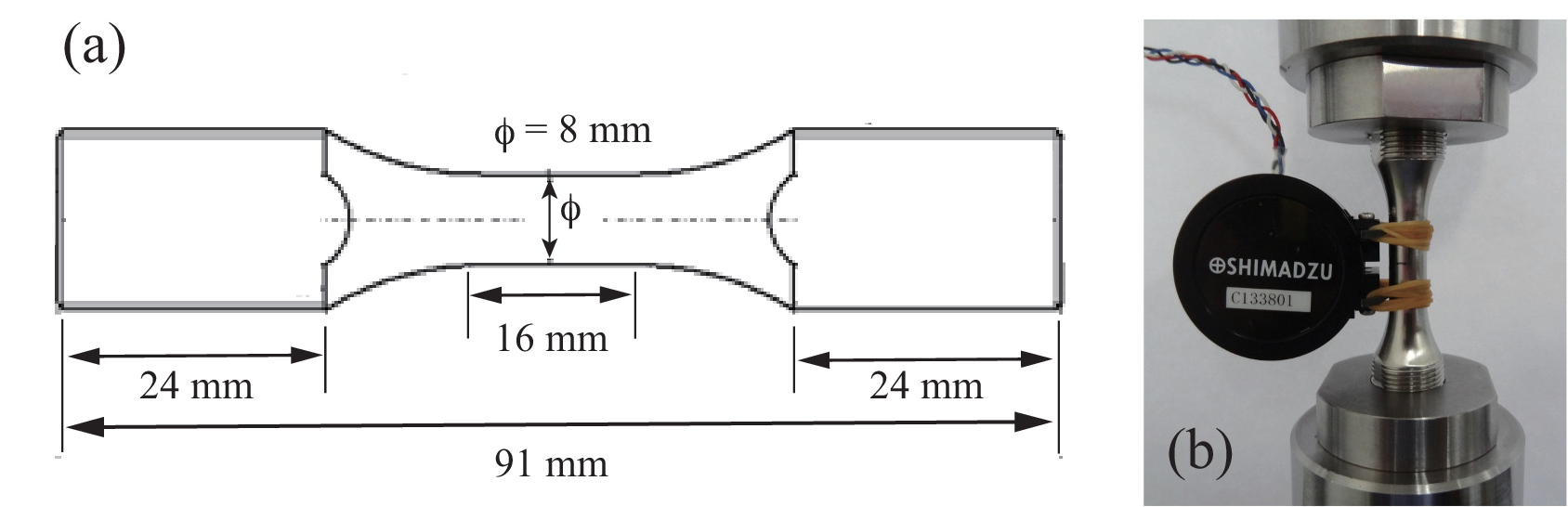}
\end{center}
\setlength\abovecaptionskip{-4pt}
\caption{(a) Dimensions of a low-cycle fatigue specimen. (b) Photograph of the fatigue test set-up} 
\label{dogbone}
\end{figure}

A 10 kg ingot of Fe-33Mn-4Si (mass$\%$) alloy was prepared by induction furnace melting in an argon atmosphere. 
The ingot was hot forged and rolled at an initial heating temperature of 1273 K into a 20 mm-thick plate. 
The plate was annealed at 1,273 K for 1 h and then quenched in water. 
The dog-bone-shaped fatigue specimen shown in Fig.\ref{dogbone}(a) was machined from the plate, and Fig.\ref{dogbone}(b) shows a photograph of the test setup. 
The axial-strain controlled tension-compression fatigue test was conducted at a total strain-amplitude of 0.01 with a triangular wave at a strain rate of 0.004 s$^{-1}$ until failure at room temperature. 
0.2 mm thick and 3 mm diameter discs were obtained from the fatigue-failed specimen using a low-speed cutter and chemical polishing in a solution consisting of hydrofluoric acid, hydrogen peroxide, and water (1:10:2 by volume). Thin foils for TEM (JEM-4000EX, JEOL; acceleration voltage of 400 kV) observations were prepared by two-step electrolytic polishing using an electrolyte composed of acetic acid and perchloric acid (10:1 by volume) during water cooling. 
\subsection{TEM observation of LPSO phase}
\begin{figure}[tb]
\begin{center}
\includegraphics[width=0.95\linewidth]{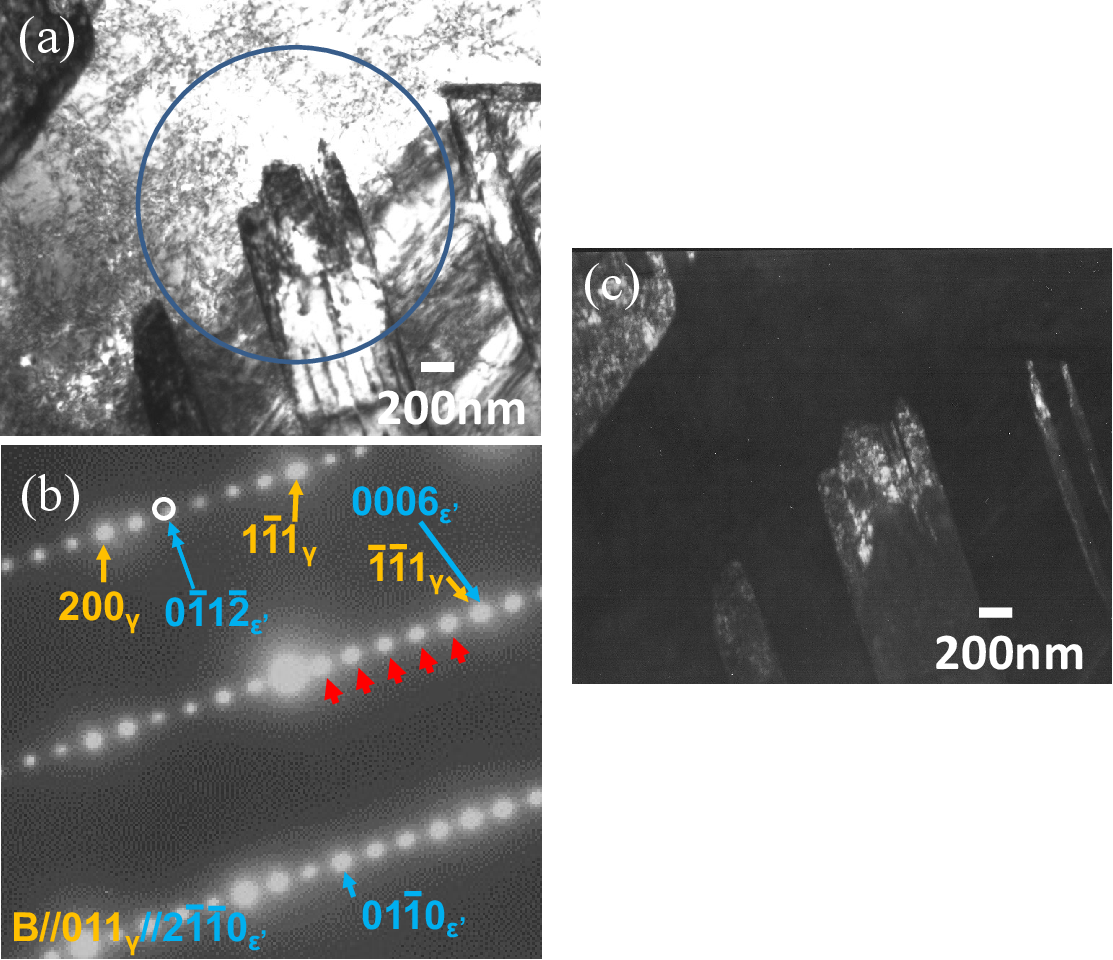}
\end{center}
\setlength\abovecaptionskip{-4pt}
\caption{ (a) Bright-field image and (b) SAED pattern of the LPSO--like ($\epsilon^{\prime}$) phase found in the fatigue-failed Fe-33Mn-4Si (mass\%) alloy (encircled area in (a)). 
The red arrows indicate the extra spots at the 1/6 positions that correspond to the ($\bar{1}$$\bar{1}$1)$_{\gamma}$ plane [the 1/3 positions corresponding to the $\{$0002$\}$ plane of the $\epsilon$ phase], the latter of which corresponds to the (0006)$\epsilon^\prime$.
(c) Dark-field image of (a). 
}
\label{TEM_LPSO}
\end{figure}

Here, we present experimental evidence for the presence of an LPSO-like structure in the Fe-Mn-Si alloys after being subjected to low-cycle fatigue failure in a series of studies in a search for a seismic damping alloy~\cite{sawaguchi2015design, Sawaguchi_Rev_2016, Sawaguchi2021}.
Figures~\ref{TEM_LPSO}(a)~and~\ref{TEM_LPSO}(b) show a bright-field image and a selected area electron diffraction (SAED) pattern, respectively. 
The TEM image of the Fe-33Mn-4Si (in mass$\%$) alloy was captured after the sample was subjected to cyclic tensile-compressive loading at room temperature and a constant strain amplitude of 0.01 until fatigue failure; the image depicts a fatigue failed alloy.
The SAED pattern in Fig.~\ref{TEM_LPSO}(b) was taken from the encircled area shown in Fig.~\ref{TEM_LPSO}(a). 

The $\gamma$ austenite and LPSO-like phase ($\epsilon^{\prime}$ phase) are oriented [011]$\gamma$ and [2$\bar{1}$$\bar{1}$0]$\epsilon$ to the electron beam direction. 
The red arrows in Fig.~\ref{TEM_LPSO}(b) show the periodic spots observed at the 1/6 positions between the $\{$111$\}$$\gamma$ spots, which is equivalent to the 1/3 positions that correspond to the $\{$0002$\}$ plane of the $\epsilon$ (hcp) phase. 
These spots show a hexagonal-type structure with a six-layer periodicity of a close-packed plane (6H). 
The (0006)${\epsilon^{\prime}}$ plane is parallel to the ($\bar{1}$$\bar{1}$1)$\gamma$ plane, and the extra spots are aligned in the $\mathbf{c^{*}}$ axis in the reciprocal space.
This orientational relationship between the $\gamma$ austenite and the $\epsilon^{\prime}$(6H) phases is identical to the so-called Shoji-Nishiyama (S-N) orientational relationship between the $\gamma$ and $\epsilon$ (2H) phases~\cite{nishiyama1978martensitic}. 
Figure~\ref{TEM_LPSO}(c) shows a dark-field image of the fatigue-failed alloy.
The $\epsilon^{\prime}$ phase is shown in bright contrast. 
This image was taken using the (0$\bar{1}$1$\bar{2}$)$\epsilon^{\prime}$ spot, which is encircled and indicated by the double-headed blue arrow in Fig.~\ref{TEM_LPSO}(b). 
The $\epsilon^{\prime}$ phase is highly defective, and the fringes in the bright-field image indicate stacking faults.

\begin{figure}[tb]
\begin{center}
\includegraphics[width=0.95\linewidth]{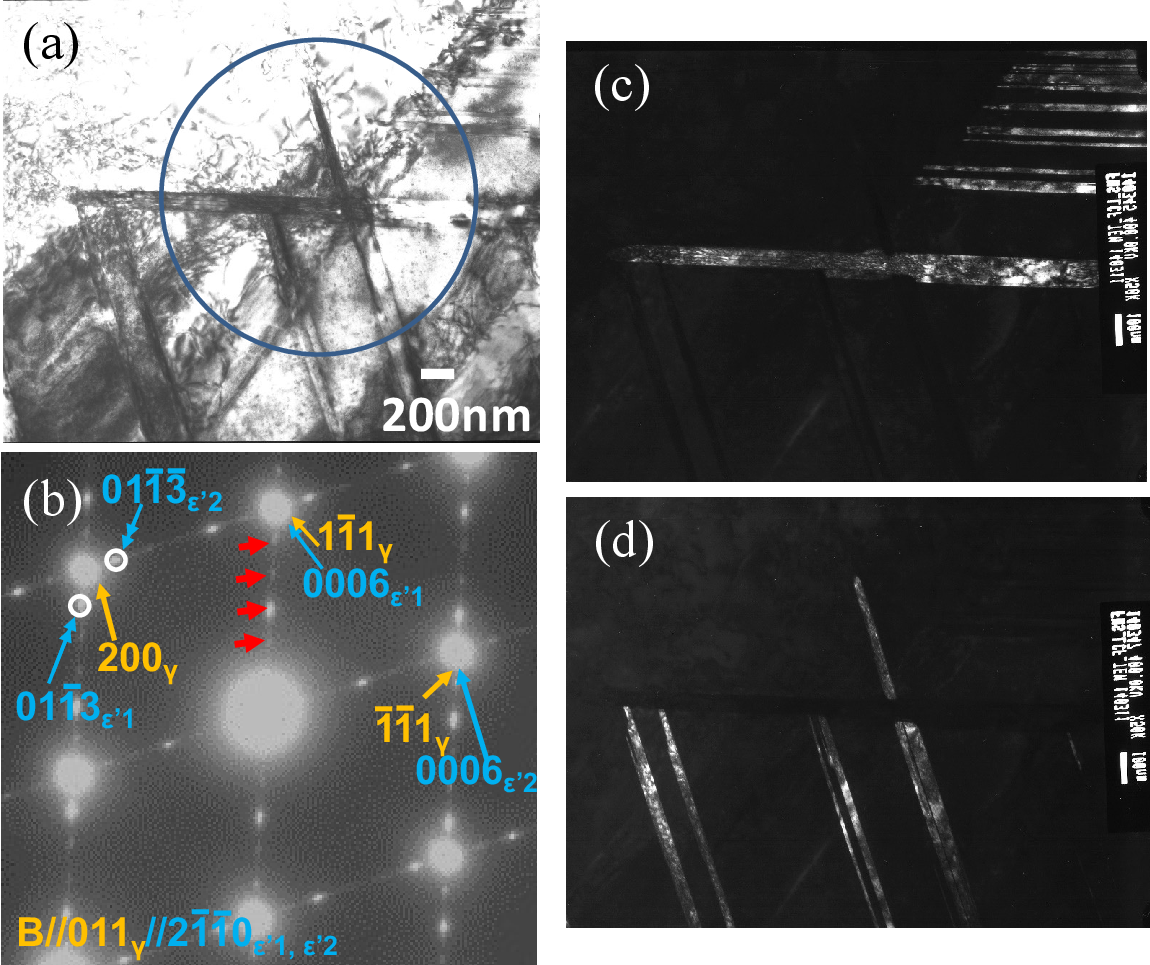}
\end{center}
\setlength\abovecaptionskip{-4pt}
\caption{
(a) Bright-field image for crossing variant plates in an LPSO-like ($\epsilon^{\prime}$) phase of the Fe-33Mn-4Si alloy, and both observed in the S-N orientation. 
(b) The (0006) planes of the variants, where $\epsilon^{\prime}_1$ and $\epsilon^{\prime}_2$ are parallel to the (1$\bar{1}$1)$\gamma$ and ($\bar{1}\bar{1}$1)$\gamma$ planes, respectively. 
The red arrows indicate the extra spots at the 1/6 positions that correspond to the (1$\bar{1}$1)$_{\gamma}$ plane (the 1/3 positions correspond to the $\{$0002$\}$ plane of the $\epsilon$ phase), the latter of which corresponds to the (0006)$\epsilon^{\prime}_1$.  
Dark-field images of the (c)~(01$\bar{1}$3)$\epsilon^{\prime}_1$ and (d)~(01$\bar{1}\bar{3}$)$\epsilon^{\prime}_2$ spots. 
The parallel plates of the $\epsilon^{\prime}_1$ and $\epsilon^{\prime}_2$ phases are shown in bright contrast in those figures. 
The streaks along the (0006)$\epsilon^{\prime}$ spots suggest a high concentration of stacking faults in the $\epsilon^{\prime}$ planes.
}
\label{TEM3}
\end{figure}
Figure~\ref{TEM3} (a) shows another TEM image for the LPSO structure taken in a different location of the specimen. 
In this area, crossing variant plates of the $\epsilon^{\prime}$ phase are observed in the S-N orientation. 
As shown in Fig.~\ref{TEM3}(b), the (0006) planes of the variants $\epsilon^{\prime}_1$ and $\epsilon^{\prime}_2$ are parallel to the (1$\bar{1}$1)$\gamma$ and ($\bar{1}\bar{1}$1)$\gamma$ planes, respectively. 
The dark-field images in Figs.~\ref{TEM3}(c) and~\ref{TEM3}(d) are taken at the (01$\bar{1}$3)$\epsilon^{\prime}_1$ and (01$\bar{1}\bar{3}$)$\epsilon^{\prime}_2$ spots, respectively. 
The parallel plates of the $\epsilon^{\prime}_1$ and $\epsilon^{\prime}_2$ phases are shown in bright contrast in the figures.
The streaks and the (0001)$\epsilon^{\prime}$ spots suggest a high concentration of stacking faults in the $\epsilon^{\prime}$ planes.
The LPSO structure is induced by cyclic deformation; therefore, the crystallographic orientation of the LPSO structure with respect to the parent $\gamma$ phase should be determined by the deformation axis. The LPSO structures shown in Figs.~\ref{TEM_LPSO}~and~\ref{TEM3} have the same six-layer periodicity and differ only in their relative orientation in the $\gamma$ matrix.

Notably, extra spots were not observed in the initial microstructure, but only in the fatigue-failed specimen after cyclic loading at room temperature. 
Unlike the Mg alloys with LSPO, the LPSO structure observed in the cyclically deformed Fe-33Mn-4Si (mass\%) alloy cannot include the ordering of solute atoms due to their relatively low diffusion rate at room temperature.
Therefore, the periodicity can only be caused by the stacking sequence of the close-packed planes, \textsf{ABC}.
The extra spots are located at the 1/3 positions that correspond to the $\{$0002$\}$ plane of the $\epsilon$ phase with the 2H structure, which indicates that the unit cell has a hexagonal (H) type structure with six-layer periodicity (6H).
It has been widely accepted that there are only two such structures without chemical ordering, 6H$_1$ and 6H$_2$, of which the stacking sequences are \textsf{ABCACB} and \textsf{ABCBCB}, respectively~\cite{Polytype1971, krishna1981close}.
However, from the TEM measurements, it is still unclear which stacking pattern of 6H is realized in the experimental LPSO phase. 

Therefore, first-principles calculations were performed to determine the structural and magnetic stabilities of the 6H structures. 
In addition, the stabilities of other structural variants, 4H and 10H, were also investigated. 
As a result, we found that the 6H$_2$ structure, which has an AFM order, was closest in energy to the hcp AFM-II structure.
Therefore, it was concluded that the $\epsilon^\prime$ phase most likely has the 6H$_2$ structure. 
The structural and magnetic properties obtained from the DFT calculations and the microscopic origin of the phase stability are reported in Sec.~\ref{CalcRes}.
It should be noted that plastic deformation modes in the deformation-induced ${\gamma}$--$\epsilon$ transformation can be found in a recent review~\cite{Sawaguchi2021}.

\begin{table}[tb]
\caption{Various structural polytypes of Fe. 
Stacking sequences are characterized by three different notations: Ramsdell, \textsf{ABC}, and \texttt{hc} notations.
The symbol $^{-}$ between layers indicates the position of a stacking fault. $\beta$ represents the hexagonality parameter.}
\label{Struct_LPSO}
\scalebox{0.9}{
\begin{tabular}{lllcccclcccrrrc}
\hline
\hline 
Type & \textsf{ABC} & \texttt{hc} \cite{christian1970dislocations} & $\beta$ (\%)\\
\hline
2H (hcp) & \textsf{AB$^{-}$} & \texttt{h} & 1.0 \\
4H (dhcp) & \textsf{ABC$^{-}$B$^{-}$} & \texttt{hc} & 0.5 \\
6H$_1$ & \textsf{ABCA$^{-}$C$^{-}$B$^{-}$} & \texttt{hcc} & 1/3 \\ 
6H$_2$ & \textsf{ABC$^{-}$BC$^{-}$B$^{-}$} & \texttt{hchhhc} & 2/3 \\
10H & \textsf{ABC$^{-}$BC$^{-}$B$^{-}$A$^{-}$CA$^{-}$C} & \texttt{cchhh} & 0.3 \\ 
3C (fcc) & \textsf{ABC} & \texttt{c} & 0 \\                 
\hline
\hline
\end{tabular}
}
\end{table}

\section{Calculation results \label{CalcRes}}
\subsection{Structural models of LPSO\label{crystals}}
We considered six different stacking sequences. 
The candidate structural polytypes are listed in Table~\ref{Struct_LPSO}.
The sequences of the 2H (hcp), 4H (dhcp), 6H$_1$, and 6H$_2$ structures are \textsf{ABAB}, \textsf{ABCB}, \textsf{ABCACB}, and \textsf{ABCBCB} stacking of close-packed layers, respectively. 
The stability of the 10H structure --an LPSO structure with a longer period than the 6H structures-- was also investigated. 
The number of possible patterns of the 10H polytype was too large to calculate all the possible structures; therefore,
an \textsf{ABCBCBACAC} stacking configuration was used, which has been observed in Mg-Zn-Y alloys~\cite{yamasaki10H}.
In addition to the \textsf{ABC} notation, we introduce a configurational notation called \texttt{hc} notation 
to characterize the different stacking variants. 
In the \texttt{hc} notation, the symbol \texttt{h} represents a local set of three layers in a hexagonal pattern (layers with identical neighbors, $i.e.$, ~\textsf{ABA} or \textsf{ACA}), and the symbol \texttt{c} represents a set of three layers in an fcc-like pattern (layers with different neighbors,~$i.e.$,~\textsf{ABC} or \textsf{BCA})~\cite{christian1970dislocations}. 
To characterize the various polytypes, the hexagonality parameter $\beta$ was employed to represent each LPSO phase as an fcc and hcp composite multilayer using \texttt{hc} notation~\cite{Raffy2002}. 
This parameter is defined as the ratio of the number of hexagonal layers ($n_h$) to the total number of layers per unit cell: $\beta$ = $n_h$/($n_h$ + $n_c$),
where $n_h$ and $n_c$ are the respective numbers of \texttt{h} and \texttt{c} blocks in each structure. 
Therefore, the $\beta$ values for the end members of the fcc and hcp structures are 0 and 1, respectively, while those of the intermediate polytypes 4H, 6H$_1$, 6H$_2$, and 10H are 1/2, 1/3, 2/3, and 0.3, respectively. Later, we discuss the ground state energies of the LPSO phases for both NM and AFM phases as a function of the $\beta$ parameter in Sec.~\ref{DFTcalc}. 

The crystal structures for non-spin-polarized calculations were first generated using periodic boundary conditions to determine their structural stabilities. 
The theoretically optimized lattice parameters and the internal atomic coordinates are listed in Tables.~\ref{Lattice_AFM_deltaE} and \ref{Coord_NM}, respectively. 
In the NM case, the hcp, dhcp, and 6H$_1$ structures of Fe belong to the same space group ($P$6$_3$/$mmc$); however, the stacking sequence along the $z$-axis of the hexagonal lattice is different. 
This stacking sequence means that the lattice constant $c$ in the hcp structure is
approximately half that of the dhcp structure.
Therefore, the unit cell of the dhcp structure contains four Fe atoms and the 6H structures contain six, while the hcp Fe structure contains two. 
Of these structures, only the 6H$_2$ structure belongs to the noncentrosymmetric $P\bar{6}m2$ space group.
Due to the limitation of symmetry operations, the 10H structure has an orthorhombic cell with the space group $Cmcm$, and it contains 20 Fe atoms (the primitive cell contains only ten atoms.).

\begin{table*}[tb]
\caption{Equilibrium lattice parameters for various structural polytypes of Fe with NM and AFM ordered states alongside optimized lattice constant ratios ($c$/$a$) in a hexagonal cell. 
Wyckoff positions and their internal atomic coordinates ($x$, $y$, $z$) for the NM and AFM states are shown in Tables~\ref{Coord_NM} and \ref{Coord_AFM}, respectively. 
The equilibrium lattice volume (zero pressure volume) $V_0$ is given per Fe atom. $B_0$ and $B^\prime$ are the bulk modulus and its pressure derivative, respectively.
$\Delta E$ is the relative total energy with respect to that of the AFM-II phase in hcp Fe. The square bracketed items of 3C indicate fixed values in the fit. $c$/$a$ is the axis ratio normalized to a 2H structure (hcp Fe).}
\label{Lattice_AFM_deltaE}
\scalebox{0.85}{
\begin{tabular}{lcccccccccccc}
\hline
\hline
Type & Ordering & Crystal &Space group & $a$ & $b$ & $c$ & $c$/$a$  & $V_0$ & $B_0$ & $B^\prime$& $\Delta E$ \\ 
        &       & system  &  & (\AA) & (\AA) & (\AA) &  & (Bohr$^3$/atom) & (GPa) &  & (meV/atom)\\
\hline 
2H~(hcp) & NM & hex & $P6_3/mmc$  & 2.46 & & 3.89  & 1.580 & 68.7 & 255 & 8.58 & 27\\ 
           & AFM-II  & ortho & $Pmcm$ & 2.47 & 4.28 & 3.98  & 1.616 & 71.0 & 199 & 4.95 & 0\\ 
           & Exp.\cite{Fei_Murphy2016}  &  &           &      & &       &       & 74.8 & 191 & 4.52 &   \\ 
           & Exp.\cite{Sakai2014}  &  &             &      & &       &       &  74.8  & 180 & 4.91 &   \\
           & Exp.\cite{YamazakiEoSFe2012} & &             &      & &       &       & 74.8 & 202 & 4.5 &   \\ 
           & Exp.\cite{Dewaele2006}  &     &        &      & &       &       &  75.8  & 165 & 4.97 &   \\  
\hline      
6H$_1$  & NM  & hex & $P6_3/mmc$ & 2.45 &   & 11.85 & 1.613 & 69.2 & 279 & 4.12 & 80 \\
        & AFM1 & ortho & $Pmnm$ & 2.47 & 4.29 & 12.02 & 1.619 & 71.7 &  175 & 5.41 & 53\\ 
        & AFM2 & ortho & $Pm2m$ & 2.47 & 4.29 & 12.04 & 1.620 & 72.0 &  183 & 7.21 & 28\\
\hline
6H$_2$ & NM   & hex & $P\bar{6}m2$ & 2.45 & & 11.78 & 1.600 & 69.0 & 285 & 5.54 & 59 \\    
       & AFM & ortho & $Pm2m$ & 2.47 &  4.29 & 11.97 & 1.611 & 71.5 & 197 & 4.50 & 15 \\            
\hline
4H & NM & hex & $P6_3/mmc$  & 2.45 & & 7.89  & 1.612 & 69.1 & 286 & 4.35 & 82 \\
10H & NM & ortho & $Cmcm$ & 2.45 & 4.25 & 19.61 & 1.600 & 69.0 & 283 & 4.43 & 60 \\ 
\hline
3C (fcc) & AFM-D & tetra & $Pmm2$ &2.50 & 2.50 & 7.08 & & 74.8 & 126 & 2.71 &56\\ 
  & AFM-S & tetra & $P4/mmm$ & 2.47 & 2.47 & 3.49 &  & 71.8 & 200 & 7.19 & 62\\ 
      & NM & cubic & $Fm\bar{3}$m & 3.45 &  &  &  &  69.2 & 279 & 4.51 & 106\\
      & Exp.\cite{Campbell_fccFe09} (293~K) &   &          &      & &       &    & [79.3]   & 133 & [5] &   \\  
      & Exp.\cite{Tsujino2013} (1273~K)    &     &        &      & &       &       &  82.7  & 111 & 5.3 &   \\  
      & Exp.\cite{DorogokupetsFe2017}    &      &       &      & &       &       &      & 146 & 4.67 &   \\  
\hline
$bcc$ & FM & cubic &$Im\bar{3}m$ & 2.83 &  &  & & 76.7 & 191 & 4.53 & $-$57\\ 
   & Exp.~\cite{Guinan1968} &             &      & &    &   &       &      & 166 & 5.29 &   \\ 
   & Exp.~\cite{DorogokupetsFe2017} &             &    &  & &       &       &      & 164 & 5.50 &   \\ 
\hline
\hline
\end{tabular}
}
\end{table*}

\subsection{Search for the antiferromagnetic order \label{strucAFM}}
\begin{figure}[tb]
\begin{center}
\includegraphics[width=0.82\linewidth]{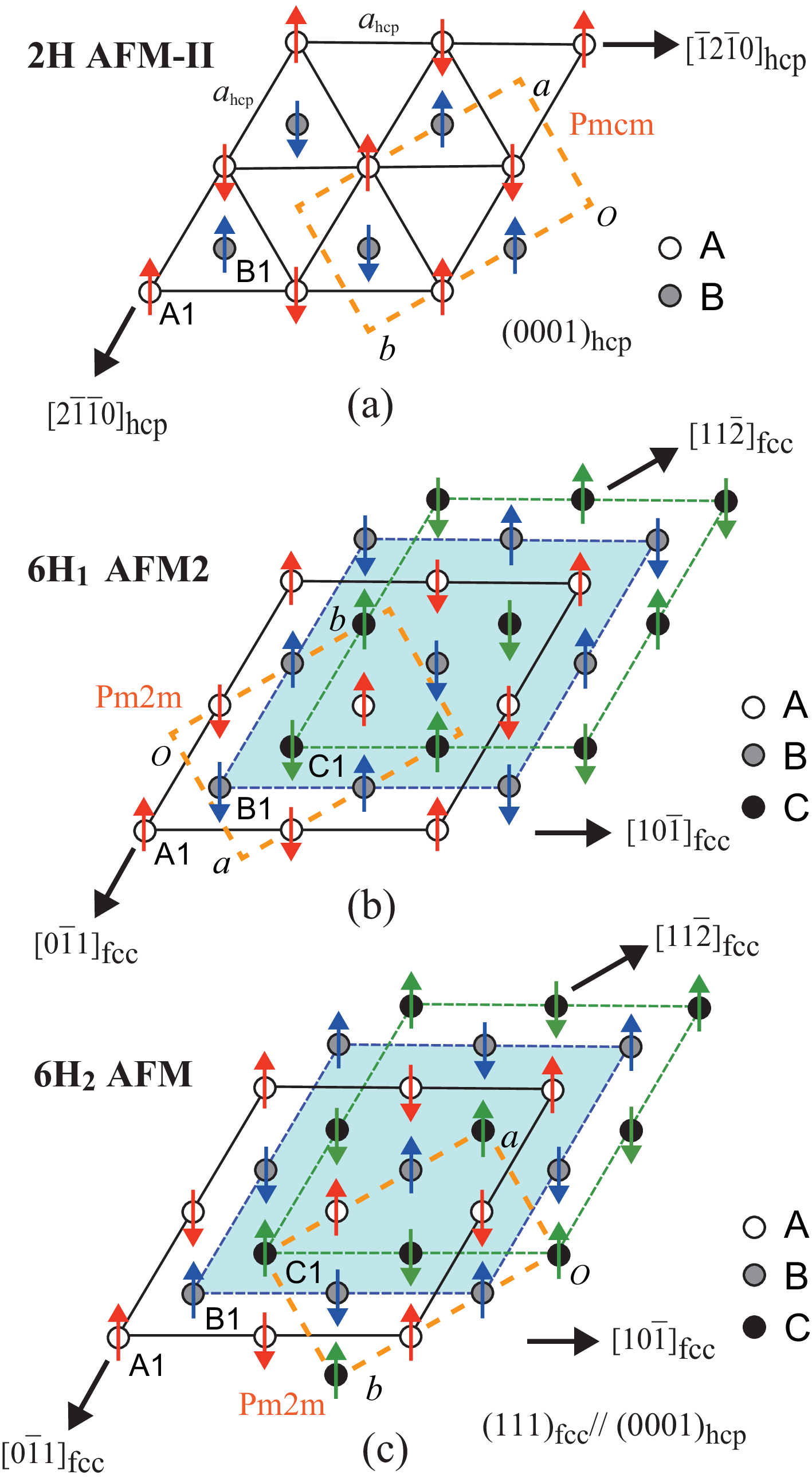}
\end{center}
\setlength\abovecaptionskip{-4pt}
\caption{(a) Projection of the hcp lattice with AFM-II order on the (0001) plane, (b) 6H$_1$--AFM2, and (c) 6H$_2$--AFM structures projected on the (111) plane of the fcc lattice, which is parallel to the (0001) plane of the hcp lattice.
The arrows denote the direction and magnitude of magnetic moments, and the magnetic moments on the \textsf{A}, \textsf{B}, and \textsf{C} layers are shown with red, blue, and green arrows, respectively. 
The orthorhombic cell is shown as bold dashed lines. 
The spin alternation along the [0$\bar{1}$1] and [10$\bar{1}$] directions of the fcc lattice is realized with this unit cell.
The \textsf{A} layer is on top of the \textsf{B} layer, and the \textsf{C} layer is on top of the \textsf{B} layer. 
The parallelogram shown in the background is the \textsf{B} layer.
The planes shown in (b) and (c) correspond to the bottom three layers of the orthorhombic cells shown in Figs.~\ref{AFM_LPSO}(a) and~\ref{AFM_LPSO}(b), respectively. 
} 
\label{hexa_ortho}
\end{figure}

The stability of the magnetic structure in LPSO phases was also computed from first-principles. 
Previous DFT studies of hcp Fe have proposed two collinear AFM configurations~\cite{Steinle-Neumann_Fe99, Steinle-Neumann_Fe_Eratum}.  
First, in the AFM-I configuration, ferromagnetically ordered basal planes of hcp lattice alternate the spin direction. 
Second, in the AFM-II configuration, the spins alternate along the hcp lattice $a$-axis, as shown in Fig.~\ref{hexa_ortho}(a).
The latter spin configuration is more energetically favorable than the former~\cite{Steinle-Neumann_Fe99, Steinle-Neumann_Fe_Eratum, Lizarraga_Fe_PRB08}.  

In order to identify stable AFM states for the 6H$_1$ and 6H$_2$ structures,
we first use the same spin alternation pattern as the AFM-II [Fig.~\ref{hexa_ortho}(a)] structure in hcp Fe for all (0001)$_{hcp}$ planes (either of type \textsf{A}, \textsf{B}, or \textsf{C}), which provides an in-plane AFM order and offers two possibilities for spin arrangements on this plane (deduced from each other by exchanging up-spins and down-spins).
In the framework of the spin-polarized DFT calculations for collinear AFM order, the atomic coordinates with different spins were designated as crystallographically independent sites. 
Therefore, the in-plane AFM patterns of the hcp lattice must be described by an orthorhombic representation of the hexagonal unit cell~\cite{Steinle-Neumann33}.
Therefore, for each six-plane LPSO structure, there are 64 (= 2$^6$) possible combinations of AFM orders to arrange two types of spin in six basal planes with the in-plane AFM arrangements. 
Of the 64 possible configurations, the patterns with opposite spin signs were equivalent, so that we obtain 32 patterns per the 6H structure as the possible initial spin configurations.
32 orthorhombic structures containing 12 atoms were thus prepared and their total energies were minimized by spin-polarized DFT calculations.
As a result, five and three types of spin configurations were determined for the 6H$_1$ and 6H$_2$ structures, respectively. The magnetic moments of all other initial spin configurations were relaxed to zero by the energy minimization process.
Furthermore, to clarify which structural polytypes of iron could be taken as an intermediate phase between the fcc and hcp phases, the total energy of fcc Fe (AFM-D, which is defined in the final paragraph of this subsection) was selected as a criterion to select stable AFM order from the eight patterns obtained.
As a result, two types of spin configurations were determined for the 6H$_1$ structure; these AFM phases are referred to as type 1 (AFM1) and type 2 (AFM2), respectively. 
The space groups of orthorhombic cells for AFM1 and AFM2 are $Pmnm$ and $Pm2m$, respectively, where the latter spin order configuration is more stable than the former. 
Therefore, the magnetic structure of the AFM2 phase is mainly discussed below. 
A stable AFM order was also identified in the 6H$_2$ structure.
This AFM ordered 6H$_2$ phase has lower energy than the 6H$_1$ structures, and it is more energetically close to the most stable structure of hcp Fe. 

The space group of the magnetic unit cell is identified from the combination of the symmetry operations in this system.  
The symmetry operations that describe the symmetry group of the orthorhombic structures were searched for using the FLAPW method.
In this DFT method, symmetry operations are applied for the internal coordinates to reduce the number of computations required. 
The symmetry operations are associated with the general positions ($x$, $y$, $z$) of the space group. 
The Wyckoff (general) positions for the atomic coordinates of the magnetic unit cells were then identified using the international tables of crystallography~\cite{Int_Table}.

The orthorhombic cell represented a hexagonal unit cell with an in-plane AFM pattern of the hcp lattice. 
Therefore, the $b$/$a$ axis ratio was fixed at $\sqrt{3}$ when optimizing the lattice constants of the orthorhombic structures. 
The $c$/$a$ ratio was optimized by fixing the lattice volume at equilibrium.
The orthorhombic cells are generated by transforming crystal axes from the parent hexagonal structure by specifying the 3$\times$3 rotation matrix shown in Appendix~\ref{AppA} and then moving the origin according to the spin configuration. 
The lattice constants are taken as $a$ = $a_{h}$,~$b$ = $\sqrt{3}$$a_{h}$, and $c$ = $c_{h}$.
The optimized lattice parameters and atomic coordinates are summarized in Tables~\ref{Lattice_AFM_deltaE} and \ref{Coord_AFM}, respectively. 
It should be noted that the orthorhombic cell for the hcp structure in this study is equivalent to the previously reported structure for AFM-II~\cite{Steinle-Neumann_Fe99, Steinle-Neumann_Fe_Eratum,Steinle-Neumann33}.

\begin{table}[tb]
\centering
\caption{Internal atomic coordinates ($x$, $y$, $z$) of NM states for various structural polytypes of Fe. 
Site represents the Wyckoff positions. The lattice constants that correspond to atomic coordinates are listed in Table~\ref{Lattice_AFM_deltaE}.}
\centering
\label{Coord_NM}
\begin{tabular}{lccccc}
\hline
\hline 
Type & Space group & Site &  & Atomic coordinates  & \\
         &  &    & $x$ & $y$  & $z$ \\
\hline
2H & $P6_3/mmc$ & $2c$ & 1/3 & 2/3 & 1/4 \\
4H  & $P6_3/mmc$ & $2a$ & 0 & 0 & 0 \\
6H$_1$ & $P6_3/mmc$ & $2b$ & 0 & 0 & 1/4 \\
       & &  $4f$ & 1/3 & 2/3 & --0.0863\\
6H$_2$ & $P\bar{6}m2$ & $1c$ & 1/3 & 2/3 & 0 \\
      &  & $2i$ & 2/3 & 1/3 & 0.166 \\
      &  & $2g$~& 0 & 0 & 1/3  \\
      &  & $1f$~& 2/3 & 1/3 & 1/2 \\
10H    & $Cmcm$ &  $4c$ & 0 & 0.6666 & 1/4 \\   
           &                &  $8f$ & 0 & 0 & 0.8489  \\        
           &                & $8f$ & 0 & 0.3330 & --0.0515  \\   
3C    & $Fm\bar{3}$m & 4$a$ & 0 & 0 & 0 \\                 
\hline
\hline
\end{tabular}
\end{table}

\begin{figure}[tb]
\begin{center}
\includegraphics[width=0.85\linewidth]{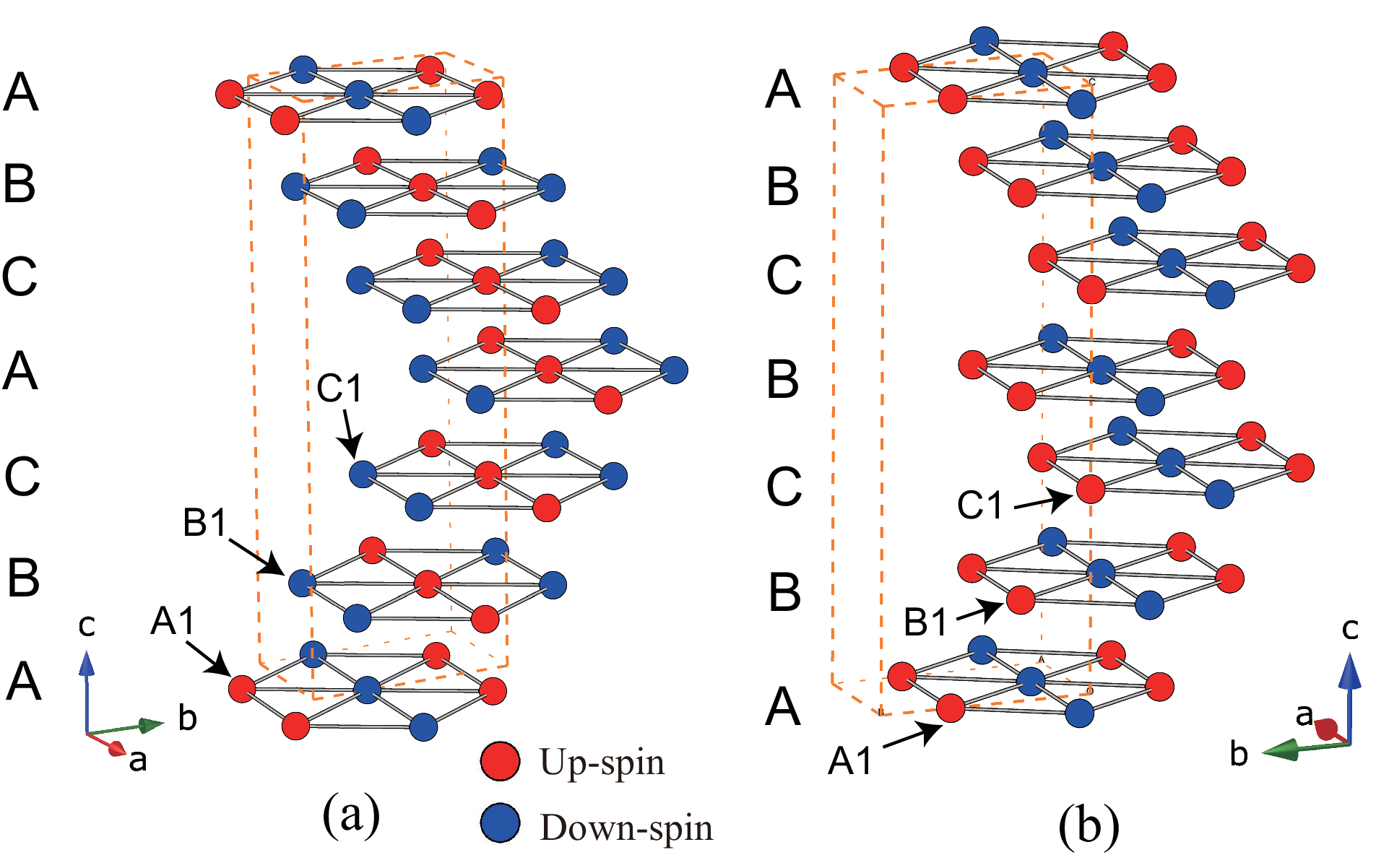}
\end{center}
\setlength\abovecaptionskip{-4pt}
\caption{Schematic illustrations of the spin arrangements for the (a) 6H$_1$--AFM2 and (b) 6H$_2$--AFM structures with the spin direction (red and blue circles represent Fe atoms with up and down spins, respectively). Orthorhombic cells are shown as bold dashed lines. 
The $ab$ planes in (a) and (b) corresponds to the \{111\} plane of the $\gamma$ phase. 
The relative position of hexagonal units due to Shockley partial dislocations are displayed;
In (a) and (b), the lattice vector $b$ corresponds to the direction of the Shockley partial dislocation along [11$\bar{2}$] of the fcc structure. 
}
\setlength\abovecaptionskip{0pt}
\label{AFM_LPSO}
\end{figure}

Figures~\ref{hexa_ortho}(b) and \ref{hexa_ortho}(c) show the spin arrangements on the $ab$ plane of the orthorhombic cells in the 6H$_1$ and 6H$_2$ structures, respectively. 
This plane is equivalent to the (111) plane of fcc, which is parallel to the (0001) plane in the hexagonal lattice. 
Only the bottom three layers (\textsf{ABC}) of the magnetic unit cell shown in Figs.~\ref{AFM_LPSO}(a) and~\ref{AFM_LPSO}(b) are depicted in these figures to analyze the difference in the spin alignment between adjacent planes. 
As shown in Fig.~\ref{hexa_ortho}, a specified Fe atom in each plane is assigned as \textsf{A1}, \textsf{B1}, and \textsf{C1}. 
When the \textsf{A1} site has up-spin in the 6H$_1$ structure [Fig.~\ref{hexa_ortho}(b)], the magnetic moments on the \textsf{B1} and \textsf{C1} sites have down-spins.
On the one hand, the magnetic moments at the \textsf{A1}, \textsf{B1}, and \textsf{C1} sites in the 6H$_2$ structure [Fig.~\ref{hexa_ortho}(c)] have the same spin orientation.

In both 6H AFM unit cells, the calculated magnetic moments on the four crystallographically inequivalent atoms for each type of spin are all different from each other.
Moreover, two atoms in the same (0001) plane (the same $z$--position of atomic coordinates in Table~\ref{Coord_AFM}) have the same magnetic moment with an opposite direction of spin.

\begin{table}[tb]
\centering
\caption{Atomic coordinates ($x$, $y$, $z$) of AFM states for various structural polytypes of Fe. 
$m_{spin}$ represents spin magnetic moments per atom within the muffin-tin sphere. $\sigma$ represents the spin index, $\uparrow$ or $\downarrow$. 
Site represents the Wyckoff positions. The lattice constants that correspond to the atomic coordinates are listed in Table~\ref{Lattice_AFM_deltaE}.}
\centering
\label{Coord_AFM}
\scalebox{0.85}{
\begin{tabular}{lclccccr}
\hline
\hline
Type & Ordering & $\sigma$ & Site &  & Atomic coordinates & & $m_{spin}$\\
         & (Space group) &      &           & $x$ & $y$ & $z$ & ($\mu_B$)\\
\hline
2H & AFM-II & $\uparrow$ & $2c$ & 0 & 0.3333 & 1/4 & 1.17\\
   & ($Pmcm$) & $\downarrow$ & $2c$ & 1/2 & 0.8333 & 1/4 & 1.17 \\
\hline
6H$_1$ & AFM1 & $\uparrow$ & $2a$ & 1/4  & --0.2407 & 1/4 & 0.95 \\ 
       & ($Pmnm$) & $\uparrow$ & $4f$ & 1/4 & 0.092 &  0.4138 & 1.47 \\
       & & $\downarrow$ & $2b$ & --1/4 & 0.2593 & 1/4 & 0.95\\
       & & $\downarrow$ & $4f$ & 1/4 & 0.4081 & --0.4138 & 1.47 \\  
       & AFM2 & $\uparrow$ & 1$a$ & 0 & 0.2757 & 0 & 1.33 \\
       & ($Pm2m$) & $\uparrow$~& 1$d$ & 1/2 & 0.7244 & 1/2 & 1.19 \\
       & & $\uparrow$~& 2$h$ & 1/2 & 0.4088 & 0.6650 & 1.40 \\
       & & $\uparrow$~& 2$h$ & 1/2 & 0.0912 & 0.8350 & 1.38 \\
       & & $\downarrow$ & 1$c$ & 1/2 & 0.7757 & 0 & 1.33\\
       & & $\downarrow$~& 1$b$ & 0 & 0.2244 & 1/2 & 1.19  \\
       & & $\downarrow$~& 2$g$ & 0 & --0.0912 & 0.6650 & 1.40   \\
       & & $\downarrow$~& 2$g$ & 0 & 0.5912 & 0.8350 & 1.38 \\
\hline
6H$_2$ & AFM & $\uparrow$ & 1$d$ & 1/2 &   0.1492 &   0  &  1.39 \\  
       & ($Pm2m$) & $\uparrow$~& 1$c$ & 1/2 & --0.1657 &  1/2 &  1.09 \\   
       &  & $\uparrow$~& 2$h$ & 1/2 & --0.4904 & --0.3359 & 1.28 \\  
       &  & $\uparrow$~& 2$h$ & 1/2 & --0.1681 & --0.1671 & 1.38 \\  
       &      &  $\downarrow$ & 1$b$ & 0 & --0.3508 &   0 & 1.39 \\ 
       &      &  $\downarrow$ & 1$a$ & 0 &   0.3343 &   1/2 & 1.09 \\  
       &      &  $\downarrow$ & 2$g$ & 0 &   0.0096 & --0.3359 & 1.28 \\   
       &      &  $\downarrow$ & 2$g$ & 0 &   0.3320 & --0.1671 & 1.38 \\  
\hline
3C    & AFM-S & $\uparrow$~& 1$a$ & 0 & 0 & 0 & 1.36 \\  
      & ($P4/mmm$) & $\downarrow$ &1$d$ & 1/2 & 1/2 & 1/2 & 1.36\\
      & AFM-D & $\uparrow$~& 1$a$ & 0   & 0   & 0   & 1.48 \\  
      & ($Pmm2$) & $\uparrow$ & 1$d$ & 1/2 & 1/2 & $-$0.25 & 1.48 \\
      &       & $\downarrow$~& 1$a$ & 0   & 0   & 0.5 & 1.48 \\
      &       & $\downarrow$~&1$d$ & 1/2 & 1/2 & 0.25 & 1.48 \\
\hline
\hline
\end{tabular}
}
\end{table}
The spin intensity of the 6H AFM structures is modulated along the $c$ axis, as listed in Table~\ref{Coord_AFM}. The magnitude of the magnetic moment (0.95 $\mu_B$) in the 2$a$ (2$b$) site of the 6H$_1$ structure in the AFM1 state is much smaller than that of the 4$f$ site (1.45 $\mu_B$).  
The results show that the decrease in the magnetic moment is favored by the geometrical frustration and the large spin degeneracy of Fe sites in the tetrahedral geometry.
In addition, the frustrated spin state of AFM1 can change to the more stable spin order of AFM2, which suggests the effect of spin frustration.    

Figure~\ref{AFM_LPSO} depicts the relative position of the close-packed layers due to the Shockley partial dislocations in the 6H structures.
The partial dislocations occur along the $b$ axis of the orthorhombic cells, parallel to the [11$\bar{2}$] direction of the fcc lattice.
An attractive feature is identified in the 6H$_1$--AFM2 structure, in that the direction of the magnetic moment is reversed only in the \textsf{A} layer located at the origin of the $c$ axis, as shown in Fig.~\ref{AFM_LPSO}(a), which indicates that the spin structure has a six-fold period with respect to the Shockley partial dislocation. 
The change in the direction of the magnetic moments corresponds to the up-spin of the \textsf{A1} site, while the \textsf{B1} and \textsf{C1} sites have down-spins, as shown in Fig.~\ref{hexa_ortho}(b).
The direction of spins does not change for the 6H$_2$--AFM structure with the partial dislocation, as depicted in Fig.~\ref{AFM_LPSO}(b); the magnetic moments at the \textsf{A1}, \textsf{B1}, and \textsf{C1} sites have the same spin orientation, as shown in Fig.~\ref{hexa_ortho}(c). 
However, in the stacking sequence of atomic layers, the \textsf{A} layer appears every six layers, and the other layers are stacked in a \textsf{BCBCB} pattern.

\begin{figure}[tb]
\begin{center}
\includegraphics[width=0.8\linewidth]{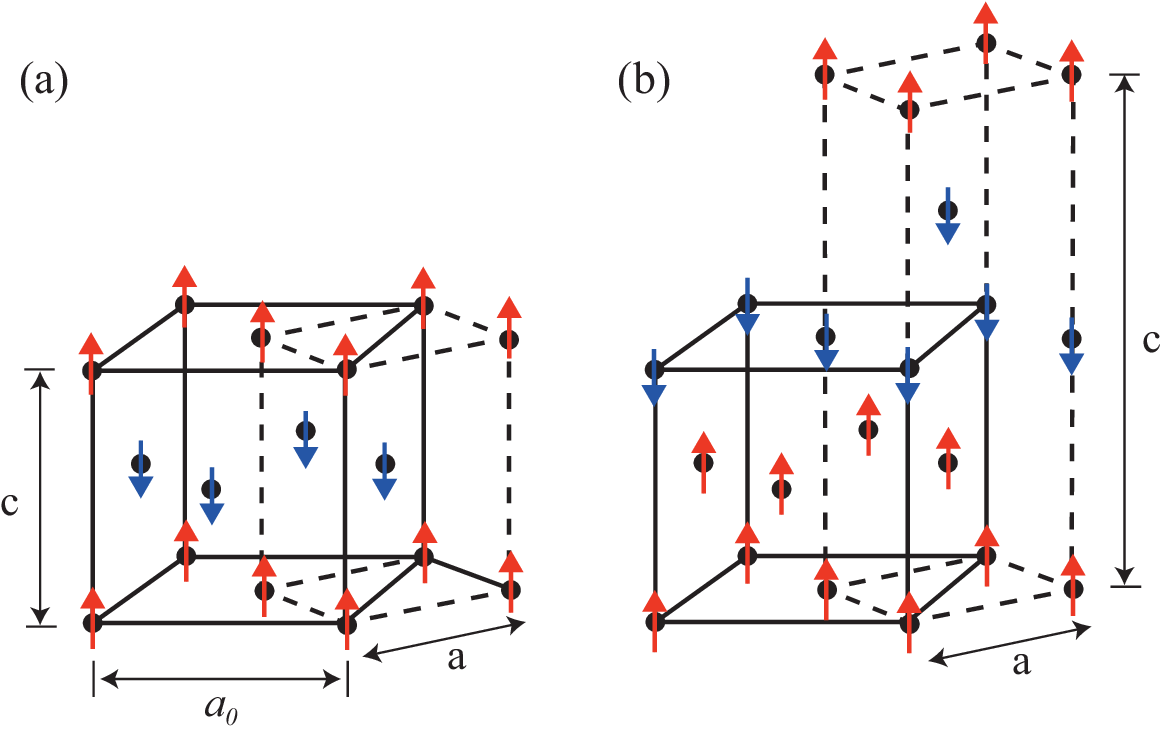}
\end{center}
\setlength\abovecaptionskip{-4pt}
\caption{
Body-centered tetragonal ($bct$) cells for the (001)-type AFM order structures of fcc--Fe;
(a) AFM single-layered (AFM-S) and (b) AFM double-layered (AFM-D) structures.
Black spheres with red up-arrows represent Fe atoms with up-spin, and those with blue down-arrows represent those with down-spin. 
The lattice constants in the $bct$ cell were used; (a) as $a$ = $a_0\sqrt{2}$ and $c$ = $a_0$ [= 2$a_0$ in (b)], where $a_0$ is the cubic lattice constant. 
Solid and dashed lines indicate the original cell with an fcc lattice and the magnetic unit cell with the $bct$ lattice, respectively. 
The $bct$ cell contains two and four atoms to describe the (001) layered structure of AFM-S and AFM-D, respectively. 
}
\label{AFM_BCT}
\end{figure}

For the fcc structure, two possible AFM states have been reported previously using body-centered tetragonal ($bct$) lattices~\cite{Herper1999_Fe}. 
The relationship between fcc and $bct$ cells with (001) type AFM spin patterns is shown in Fig.~\ref{AFM_BCT}. 
In both spin arrangements, the magnetic moments are parallel to each other in the (001) plane. 
However, the first arrangement, termed the AFM single layer (AFM-S), has alternating layers of spin-up and spin-down along the [001] axis. In the second arrangement, double layers with ferromagnetic interlayer coupling are AFM ordered along the [001] direction; therefore, this arrangement is termed an AFM double layer (AFM-D). 
The magnetic unit cells of AFM-S and AFM-D belong to the space groups $P4/mmm$ and $Pmm2$, respectively. 
The $c$/$a$ ratios for AFM-S and AFM-D are $\sqrt{2}$ and $2\sqrt{2}$, respectively, which corresponds to the fcc structure, and these values are fixed during the structural optimization. 
These calculations verify the result that the latter AFM pattern is more energetically favorable than the former.

\subsection{Structural and magnetic phase stability\label{DFTcalc}}
In this section, we discuss the results for the structural and magnetic phase stability of LPSO structures.
Table~\ref{Lattice_AFM_deltaE} shows the optimized structural parameters and the total energy differences with respect to the AFM-II state of hcp Fe. 
Among the NM states of the LPSO candidates, the 6H$_2$ and 10H structures are energetically close to each other; however, the 6H$_2$ structure is the most energetically close to hcp Fe. This result for the NM states is similar to the results of DFT studies for pure Mg, in which the energy difference between various LPSO phases is quite small~\cite{Iikubo_alpha_Mg}. 
On the other hand, the 6H$_1$ structure is less energetically favorable than 6H$_2$, although energetically close to dhcp Fe.

\begin{figure}[tb]
\begin{center}
\includegraphics[width=0.95\linewidth]{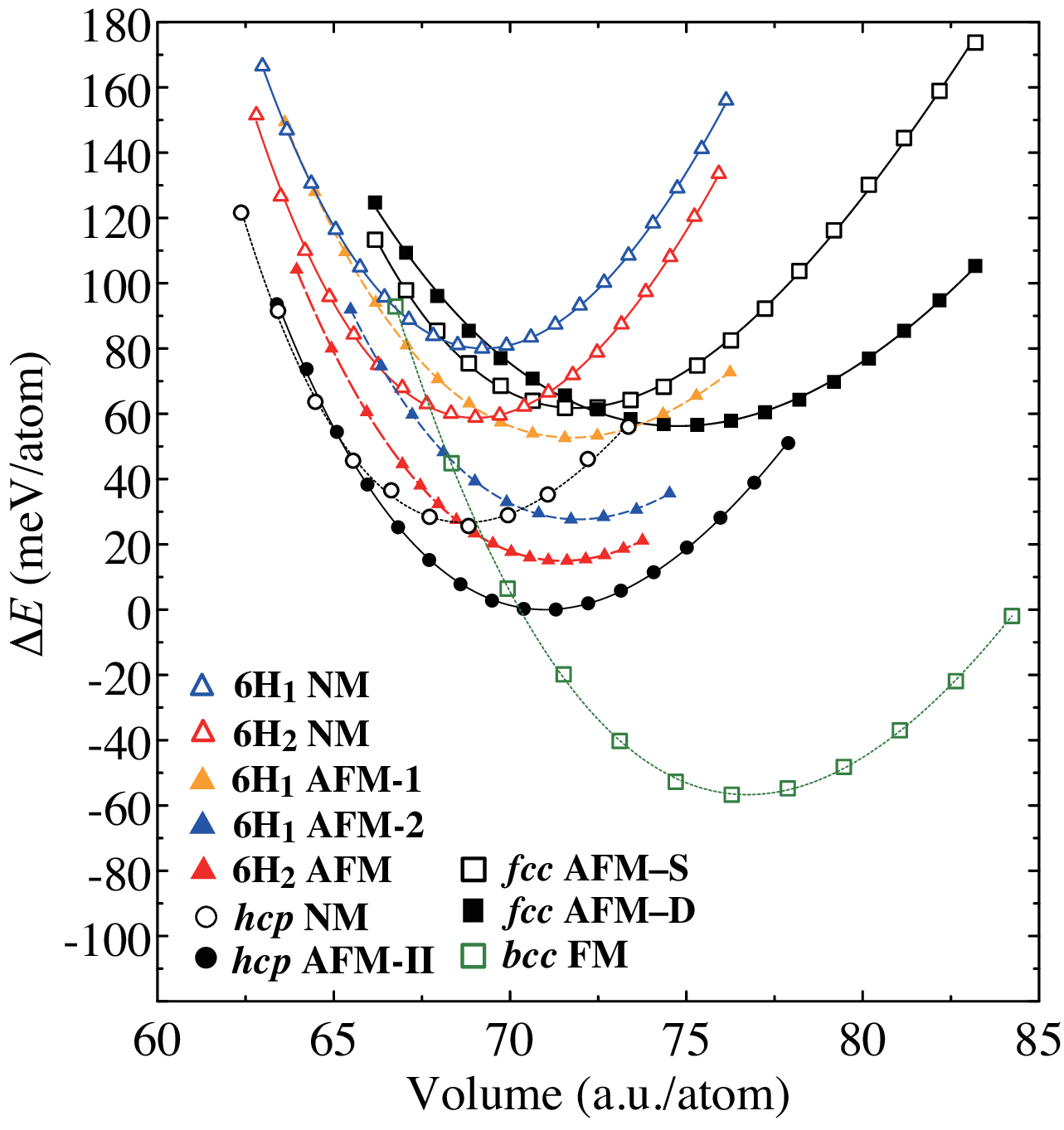}
\end{center}
\setlength\abovecaptionskip{-4pt}
\caption{
Volume dependence of the total energy difference in structural polytypes of Fe with respect to the hcp AFM-II phase.
The total energies of each phase are calculated as a function of the lattice volume per atom within the GGA-PBE functional. 
Open and solid squares on black lines represent the AFM-S and AFM-D states in the fcc structure. 
Solid orange and blue triangles on dashed lines represent the AFM-1 and AFM-2 states for the 6H$_1$ structures, respectively. 
Open and solid triangles on red dashed lines show NM and AFM states for the 6H$_1$ structures, respectively.
Open squares show the fcc structure with the AFM-D ordered state.  
Open and solid circles show NM and AFM-II states of the 2H (hcp) structures, respectively.
Open green squares on the green dotted line represent the ferromagnetic (FM) state of $bcc$ Fe.
}
\setlength\abovecaptionskip{0pt}
\label{TEN_AFM_LPSO}
\end{figure}

\begin{figure}[tb]
\begin{center}
\includegraphics[width=0.8\linewidth]{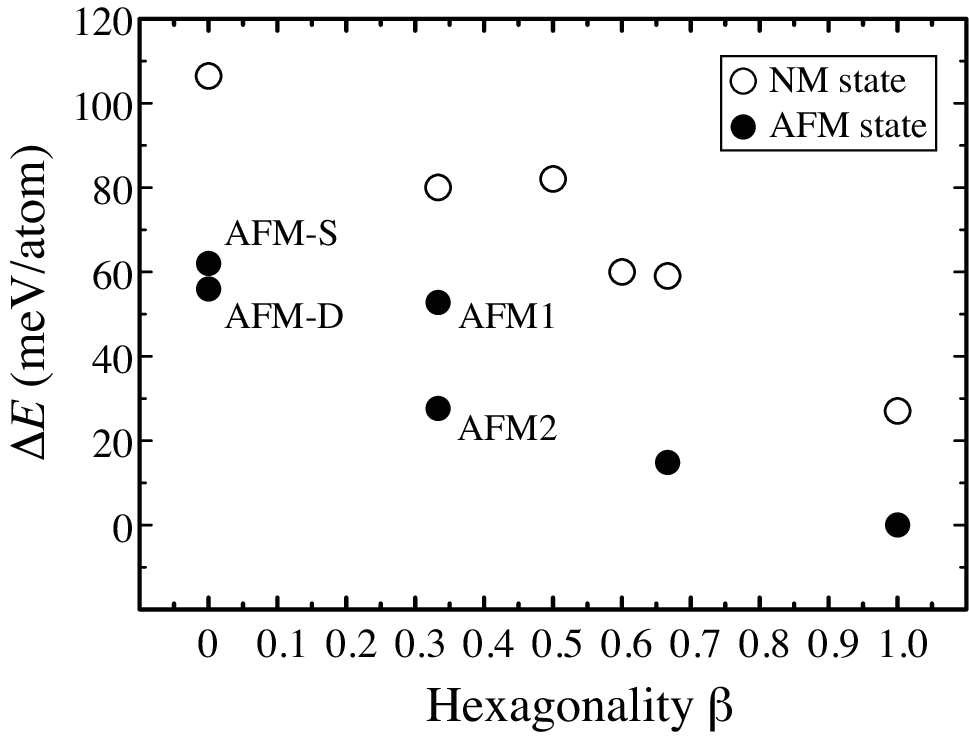}
\end{center}
\setlength\abovecaptionskip{-4pt}
\caption{Difference of the total energies for different structural polytypes of pure iron as a function of hexagonality, $\beta$. 
$\Delta$$E$ was calculated with respect to hcp Fe with the AFM-II phase and are listed in Table~\ref{Lattice_AFM_deltaE}. 
Open and solid circles represent NM and AFM states, respectively. 
The values of $\beta$ are listed in Table~\ref{Struct_LPSO}.
}
\label{ConvexHull}
\end{figure}

FLAPW calculations with the spin-polarized form of the GGA-PBE functional were also performed, and stable AFM spin structures were determined for the 6H$_1$, 6H$_2$, hcp, and fcc structures (Table~\ref{Lattice_AFM_deltaE}). 
Several possible magnetic order patterns were also examined for the 4H and 10H structures, including ferromagnetic and ferrimagnetic states; however, no magnetic orderings were stabilized. Therefore, Figure~\ref{TEN_AFM_LPSO} shows only the volume dependence of the total energies for 6H$_1$ and 6H$_2$ that consider AFM ordering and the NM state, as well as those for fcc and hcp Fe. 
As shown in Fig.~\ref{ConvexHull}, the AFM states of the 6H$_1$, 6H$_2$, hcp, and fcc structures are energetically lower than those of the NM phase by approximately 30--50 meV/atom. 
Figure~\ref{ConvexHull} also describes the difference of total energy ($\Delta E$) with respect to hcp Fe as a function of the hexagonality, $\beta$. 
In both the NM and AFM phases, $\Delta E$ decreases almost proportionally as $\beta$ increases. 

The equilibrium volumes and bulk moduli are also listed in Table~\ref{Lattice_AFM_deltaE}. The 6H$_2$ structure with AFM ordering is energetically closest to the ground state structure of hcp Fe at ambient pressure. While the 6H$_1$ structure is also stabilized by AFM ordering, the 6H$_1$--AFM1 structure is as unstable as fcc with AFM-D ordering. 
Nevertheless, the AFM2 state of 6H$_1$ has a lower energy than AFM1, and it becomes energetically closer to the AFM phase of the 6H$_2$ structure. There is a notable energy difference of 13 meV/atom (approximately~150~K) between the 6H$_1$--AFM2 and 6H$_2$--AFM phases. Therefore, we suggest that the LPSO-like phase observed by TEM measurements most likely has the 6H$_2$ structure.

Figure~\ref{TEN_AFM_LPSO} shows the volume dependence of the total energy difference of stacking variants with respect to the AFM-II state of hcp Fe.
The equilibrium lattice volumes for the NM states of all the structural polytypes are quite small (69 bohr$^3$/atom), and their bulk moduli (approximately 280 GPa) are generally much higher than the experimental values measured at finite temperatures. 
The corresponding values for the AFM states are slightly different from each other, which is expected to be due to the magneto--volume effect and spin--spin interactions.   
When AFM order is considered, the differences of bulk moduli between different types of LPSO become more distinct. 
As summarized in Table~\ref{Lattice_AFM_deltaE}, hcp Fe with the AFM-II state leads to a better agreement with the experimental bulk modulus within 10$\%$ overestimation~\cite{YamazakiEoSFe2012, Fei_Murphy2016, Sakai2014} than NM calculations. 
The calculated bulk modulus for $bcc$ Fe is also slightly overestimated from the experimental modulus~\cite{Guinan1968, DorogokupetsFe2017}. 
The lattice volume and bulk modulus for the AFM--D state of the fcc~(3C) structures are closer to the experimental values (6$\%$ underestimation from the experimental bulk modulus at 293~K~\cite{Campbell_fccFe09, Tsujino2013}). 
While the bulk modulus for the AFM--S state overestimates the experimental values, the equilibrium lattice volume for AFM--S agrees well with those of the hcp, 6H$_1$, and 6H$_2$ structures.  

Structural optimization for lattice parameters and internal atomic coordinates was performed for both the spin-polarized and unpolarized calculations. 
To determine the equilibrium lattice parameters in the ground state at ambient pressure, we first determined the equilibrium volume by specifying the axis ratio of $c$/$a$ at a constant value, and 
the calculated total energies at different volumes were fitted using the third-order Birch-Murnaghan EoS. 
The $c$/$a$ ratio is optimized by specifying the lattice volume at equilibrium using the fourth-order of the fitting. In Fig.~\ref{TEN_AFM_LPSO}, the energy--volume curves are plotted with the $c$/$a$ ratio optimized at ambient pressure.

\subsection{Origin of phase stabilities}
\label{origin}
It is also interesting to understand how the stacking sequence along the $c$ axis changes the electronic structure. 
The DOS for metals with an hcp structure is generally characterized by a deep valley or dip near the Fermi level ($E_F$), where $E_F$ is at the lowest position (the bottom) of the deep valley~\cite{JYamashita_BeMg73, Paxton1997, BlahaPRB_hcpMetal}.  
In the 1970s, before first-principles calculations were established, Inoue and Yamashita suggested that the depth of the valleys in the DOS represents the magnitude of splitting of the main peaks of the DOS between those in the occupied and unoccupied states~\cite{JYamashita_BeMg73}. 
This is a type of energy separation between the bonding and antibonding molecular orbitals due to the significant hybridization between the $s$ and $p$ states~\cite{Slater1965,JYamashita_BeMg73}. 
From a comparison of the electronic structure of hcp Be and hcp Mg, they also suggested that the degree of energy splitting between the two main peaks (the width of the deep valley or dip) appears as a difference in the enthalpy of formation (or cohesive energies) between the metals~\cite{JYamashita_Book73}. 
In the early 1970s, it would have been difficult to quantitatively calculate the difference in the heats of formation. 

Later, Andersen derived the force theorem, which describes how the change in the total energy of an electron system can be calculated to the first order in a virtual displacement~\cite{springford1979}.
With this theorem, the energy difference can be simply calculated with DFT as the differences of appropriate sums of the one-electron eigenvalue energies~\cite{HEINE19801, Skriver1982FT}. 
Therefore, the phase stability analysis based on DOS near the $E_F$ is effective to understand the energy change associated with small displacement based on the force theorem. 

\begin{figure}[tb]
\begin{center}
\includegraphics[width=0.95\linewidth]{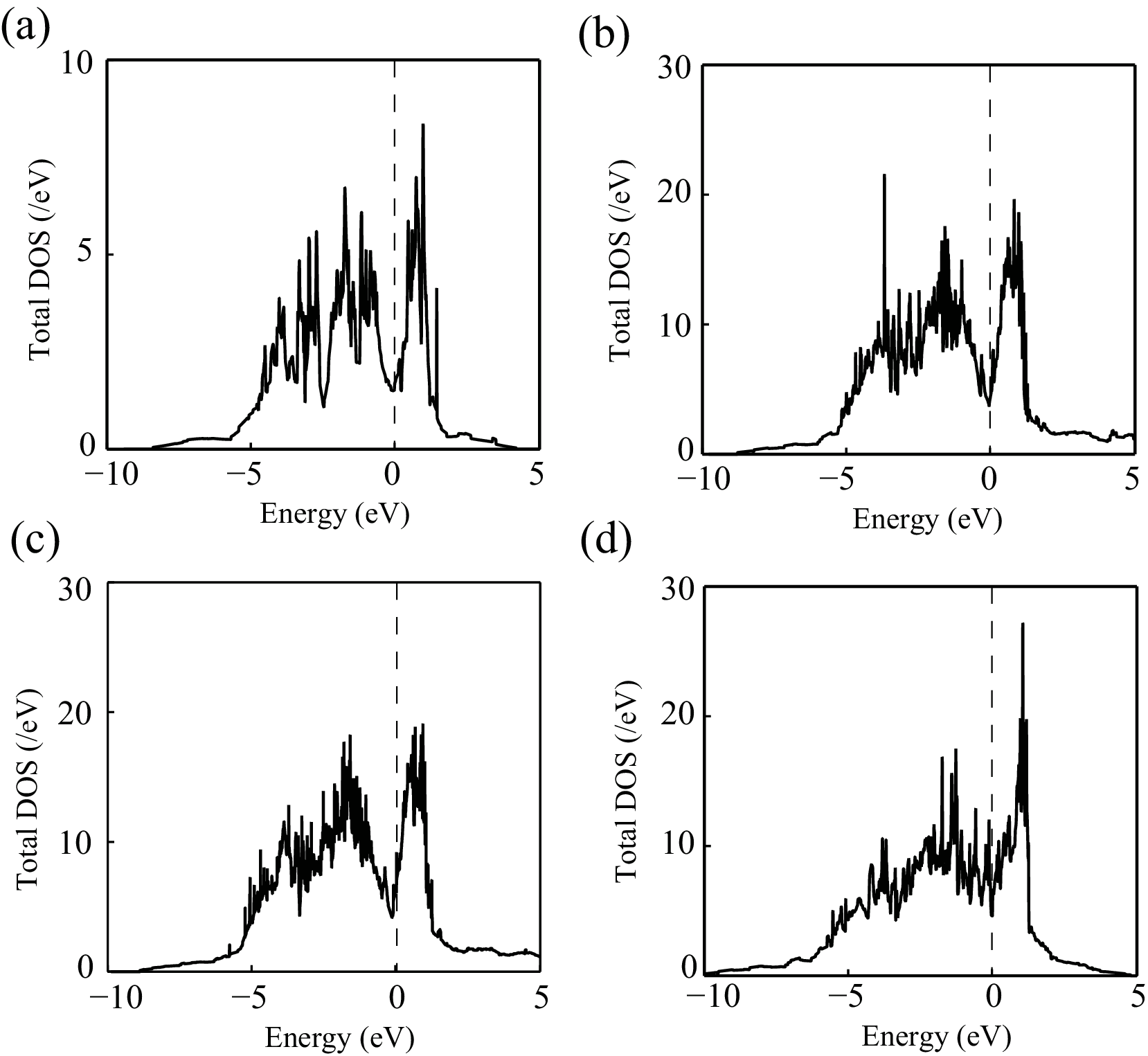}
\end{center}
\setlength\abovecaptionskip{-4pt}
\caption{The total DOS for the AFM state of the (a) 2H~(hcp), (b) 6H$_2$, (c) 6H$_1$ with AFM2 and (d) 6H$_1$ with AFM1 structures of pure iron at ambient pressure. 
Dashed lines represent the Fermi level.}
\label{DOS_AFM_LPSO}
\end{figure}
Figures~\ref{DOS_AFM_LPSO}(a)--\ref{DOS_AFM_LPSO}(d) compare the total DOS for the hcp, 6H$_2$, and 6H$_1$ structures with AFM ordering.  
A typical deep valley in the DOS observed in hcp metal is evident in Fig.~\ref{DOS_AFM_LPSO}(a). 
The total DOS for the 6H$_2$ structure also has a similar DOS valley near $E_F$~[Figs.~\ref{DOS_AFM_LPSO}(b)], and several small peaks appear near the bottom of these DOS, compared with those for hcp Fe.  
This is evident in the \textsf{ABCBCBA} stacking of the 6H$_2$ structure, where the \textsf{A} layer appears every six layers, and the other layers are stacked in a \textsf{BCBCB} pattern. 

The DOS for the 6H$_1$ structure with the AFM2 configuration is similar to that of the 6H$_2$ phase; however, the $E_F$ is located at slightly higher energy than the bottom of the valley of the DOS~[Figs.~\ref{DOS_AFM_LPSO}(c)]. 
We consider this difference in DOS to be the origin of the energy difference between the 6H$_1$ and 6H$_2$ structures. 
A comparison of the spin structures of hcp and 6H$_2$ indicates that the local spin arrangement and the stacking sequence agree well with those of the \textsf{ABAB} pattern in the hcp structure. 
Nevertheless, the total DOS for the unstable 6H$_1$ structure with AFM1 ordering has many peaks near $E_F$, and the valley of the DOS is obscured [Fig.~\ref{DOS_AFM_LPSO}(d)]. 
We have shown that this deviation of the DOS from the hcp structure is the microscopic origin of the structural stability of the candidate LPSO structures. 
Based on this analysis, one of the present authors studied the origin of phase stability in an Mg-Zn-Y alloy with LPSO, 
in which solute elements of Zn and Y were embedded in the Mg matrix near stacking faults~\cite{egusa2012structure}. 
The results will be reported elsewhere shortly. 

\section{Summary}
The structural and magnetic properties of long-period stacking order structures (polytypism) in pure iron were studied by first-principles DFT calculations. 
During deformation-induced martensitic transformation from $\gamma$-austenite to $\epsilon$-martensite, a phase (different from the $\epsilon$ phase) was discovered in Fe-Mn-Si--based alloys.
In this phase, the additional diffraction spots are located at the 1/3 positions that correspond to the $\{$0002$\}$ plane of the $\epsilon$ (hcp) phase with the 2H structure, which suggests a 6H structure.
However, the actual stacking pattern of the 6H phase is unknown.
Therefore, we proposed several structural models for the LPSO structure of pure iron, including 4H, 6H$_1$, 6H$_2$, and 10H structures, and structural optimization was performed using first-principles DFT calculations. 
From a search among the stable magnetic phases, stable AFM states were identified in the 6H$_1$ and 6H$_2$ structures. 
An AFM state of 6H$_2$ was also revealed as energetically closest to the hcp structure, 
and the observed LPSO-like phase has a high probability of adopting the 6H$_2$ structure. 
Due to the probably coherent nature between the possible 6H structures, a negligibly low elastic contribution may not affect the highest probability of the appearance of the 6H$_2$ structure.
The electronic origin of the phase stability is attributed to the depth of the valley in the DOS near the Fermi level;
 the energy splitting between the two peaks in occupied and unoccupied states is large, which maximizes the phase stability. 
The relationship between the electronic structure and the phase stability was quantitatively verified for the LPSO and hcp phases in Fe, which was proposed for hcp metals in the 1970s.  

\section{ACKNOWLEDGEMENTS}  
The authors thank A. Singh, T. Oguchi, W-T. Geng, and D. S. Shih for stimulating discussions. 
We also acknowledge D.~Drozdenko, I. Tkach, D. Korotin, and Z. Pchelkina for collecting and translating old Russian papers by Lysak $et$ $al$.
This research was funded by a Grant-in-Aid for Scientific Research (No. 19K04988, 20H00312, 21H01220, and 21H01659) from the Japan Society for the Promotion of Science (JSPS) and a CREST Grant (No. JPMJCR2094) from the Japan Science and Technology Agency (JST).
This work was performed under the GIMRT Program of the Institute for Materials Research (IMR), Tohoku University and the Cooperative Research Program of the Network Joint Research Center for Materials and Devices.
TT is supported in part by the Leading Initiative for Excellent Young Researchers (LEADER), a Ministry of Education, Culture, Sports, Science and Technology (MEXT) program, Japan. 
The computations were mainly conducted using the computer facilities of Research Institute for
Information Technology at Kyushu University, MASAMUNE at IMR, Tohoku University, and Supercomputer Center at Institute for Solid State Physics, The University of Tokyo, Japan.

\bibliography{./library}

\appendix
\section{Transformation from a hexagonal unit cell to an orthorhombic unit cell}
\label{AppA}
The crystallographic unit cell (basis) vectors of the orthorhombic cell, ${\bf{a}}_{orth}$, ${\bf{b}}_{orth}$, ${\bf{c}}_{orth}$, are related to those of the hexagonal unit cell, ${\bf{a}}_{h}$, ${\bf{b}}_{h}$, ${\bf{c}}_{h}$, by
\[
  {\bf{a}}_{orth}, {\bf{b}}_{orth}, {\bf{c}}_{orth} 
   =  ({\bf{a}}_{h}, {\bf{b}}_{h}, {\bf{c}}_{h}) 
\begin{pmatrix}
      1 & 1 & 0 \\
      -1 & 1 & 0 \\
      0 & 0 & 1
\end{pmatrix}.
\]
\end{document}